\pdfoutput=1
\documentclass[12pt,a4paper]{article}

\usepackage[utf8]{inputenc}
\usepackage[T1]{fontenc}
\usepackage{lmodern}
\usepackage[margin=1in]{geometry}
\usepackage{amsmath,amssymb}
\usepackage{booktabs}
\usepackage{siunitx}
\usepackage{graphicx}
\graphicspath{{figures/}}
\usepackage[section]{placeins}

\usepackage[numbers,sort&compress]{natbib}

\usepackage{xcolor}
\usepackage{hyperref}
\hypersetup{
  colorlinks=true,
  linkcolor=blue,
  citecolor=blue,
  urlcolor=blue
}

\begin{document}

\title{Sentiment and Volatility in Financial Markets:\\
A Review of BERT and GARCH Applications\\
during Geopolitical Crises}

\author{Domenica Mino\thanks{Corresponding author: \texttt{123110733@umail.ucc.ie}} 
\and Cillian Williamson\thanks{\texttt{CWilliamson@ucc.ie}}}
\date{\small School of Mathematical Sciences, University College Cork}

\maketitle

\begin{abstract}
Artificial intelligence techniques have increasingly been applied to understand the complex relationship between public sentiment and financial market behaviour. This study explores the relationship between the sentiment of news related to the Russia-Ukraine war and the volatility of the stock market. A comprehensive dataset of news articles from major US platforms, published between January 1 and July 17 2024, was analysed using a fine-tuned Bidirectional Encoder Representations from Transformers (BERT) model adapted for financial language. We extracted sentiment scores and applied a Generalised Autoregressive Conditional Heteroscedasticity (GARCH) model, enhanced with a Student-t distribution to capture the heavy-tailed nature of financial returns data. The results reveal a statistically significant negative relationship between negative news sentiment and market stability, suggesting that pessimistic war coverage is associated with increased volatility in the S\&P 500 index. This research demonstrates how artificial intelligence and natural language processing can be integrated with econometric modelling to assess real-time market dynamics, offering valuable tools for financial risk analysis during geopolitical crises.
\end{abstract}



% ...
\clearpage
\section{Introduction}
The influence of online news, social media, and television on public perception and investor sentiment is key to explaining the persistent fluctuations in financial markets \citep{baker2007,lo2022}. In this context, digital media has substantial influence over societal emotions, particularly in investors' sentiment during geopolitical crises such as the armed conflict between Russia and Ukraine, which began on February 22, 2022. Understanding the dynamics of this influence is crucial to analysing how stock markets respond during periods of heightened global tension. 

The literature has shown that investor sentiment significantly influences stock prices and volatility of returns \citep{engelberg2011,tetlock2007}. However, analysing investor sentiment presents challenges due to the large volume of unstructured online data, such as news articles, social media posts and online forum discussions \citep{bollen2011,sprenger2010}. Traditional sentiment analysis techniques such as lexicon-based methods and basic machine learning algorithms have encountered limitations in handling complex and diverse data embedded in financial language.

Recent advances in natural language processing (NLP), particularly through deep learning, have introduced powerful models capable of overcoming these limitations. An example is the Bidirectional Encoder Representations from Transformers model, which has significantly improved the capture of nuanced sentiment in text-based data \citep{devlin2018,liu2012}. These improvements have transformed the field of AI through architectures capable of extracting complex hierarchical representations from data \citep{schmidhuber2015}. BERT's architecture, rooted in deep neural networks, enables it to interpret complex sentence structures and semantic relationships. This makes it particularly useful for sentiment analysis in finance, where the language can be technical, context sensitive, and subtle.

This study aims to contribute to the growing knowledge of applied artificial intelligence and finance by examining how sentiment extracted from digital media influences financial markets during geopolitical instability.  Specifically, it investigates the relationship between the sentiment in U.S. headlines news coverage of the Russia-Ukraine conflict and volatility in financial markets. Given the scale and complexity of unstructured textual data in digital media, this research uses the BERT model to quantify the sentiment from news articles, while a GARCH model enhanced with a Student t distribution is used to model the volatility in market returns.  

The structure of this paper is as follows. Section 2 reviews the literature, Section 3 details the data and methodology employed in this research, Section 4 presents the results, Section 5 discusses the findings, and Section 6 offers conclusions and recommendations for future research.

\section{Literature Review and Conceptual Framework}\label{class}

This section reviews the relevant literature on the influence of investor sentiment on financial markets, particularly during geopolitical crises. It evaluates traditional and advanced sentiment analysis techniques, focusing on their application in financial contexts, and outlines how such sentiment interacts with volatility modelling through GARCH-based approaches. The review also integrates recent studies on the Russia-Ukraine conflict to contextualise the current research.

\subsection{Investor Sentiment and Financial Markets}
Investor sentiment has been extensively studied for its influence on stock market fluctuations. Sentiment-driven trading can lead to mispricing and temporary deviations from fundamental values, as documented by \citep{baker2007}. Market participants often rely on sentiment indicators to anticipate potential price movements, making sentiment analysis a critical tool in financial research. Investor emotions, such as fear and optimism, drive market trends, and understanding these psychological factors is essential for predicting asset price changes.

Previous studies have highlighted the role of media coverage in shaping investor sentiment. \citep{engelberg2011} and \citep{tetlock2007} demonstrated that financial news and media play a significant role in influencing stock returns. The widespread dissemination of news through digital platforms has further amplified the impact of media-driven sentiment on market movements. However, quantifying sentiment from unstructured textual data remains a challenge, necessitating the use of sophisticated natural language processing (NLP) techniques to capture its effects accurately.

\citep{bollen2011} and \citep{sprenger2010} explored various methodologies for measuring investor sentiment using social media and financial news. Their findings suggest that leveraging computational approaches to analyse textual data can provide meaningful insights into market behaviour. Despite advancements in sentiment quantification, limitations persist due to the complexity and context-dependent nature of financial texts, necessitating further research in this domain.

\subsection{Limitations of Traditional Sentiment Analysis Methods}
Lexicon-based sentiment analysis methods, while accessible and interpretable, are often too simplistic for financial contexts. These texts often involve domain-specific jargon, ambiguous phrasing, and implicit sentiment, making them difficult to interpret using standard techniques \citep{lo2022,loughran2011}. Hence they can misclassify sentiment in nuanced financial statements, especially when domain-specific language is involved.

Machine learning models, including support vector machines (SVM) and naive Bayes classifiers, have been employed to improve sentiment classification accuracy. However, these models often require extensive labelled datasets for training, and their performance may degrade when applied to new or evolving financial texts \citep{pang2008}. On the other hand, Popular tools like VADER \citep{hutto2014} have demonstrated improvements in handling informal texts, but still fall short when interpreting ambiguous financial expressions or negation-heavy news articles. 

These limitations reduce their predictive accuracy in high-stakes domains such as financial forecasting. Sentiment in a financial context is frequently influenced by implicit meanings, making it challenging to develop generalised models that perform well on different data sets. With the rise of social media platforms and real-time news dissemination, sentiment analysis models must adapt to the rapid influx of unstructured textual data \citep{bollen2011}. The dynamic nature of market sentiment and the evolving language used in financial discussions underscores the need for more robust and context-aware NLP techniques. Advanced deep learning models offer promising solutions to these challenges by leveraging neural networks to better understand sentiment dynamics \citep{schmidhuber2015}.

\subsection{Deep Learning and Transformer Models}
The development of deep learning and transformer-based models has revolutionised sentiment analysis. One of the most notable advances is the Bidirectional Encoder Representations from Transformers (BERT) model, which improves contextual understanding by leveraging bidirectional training \citep{devlin2018}. Unlike traditional sentiment analysis techniques that rely on predefined lexicons or basic machine learning models, BERT can dynamically capture semantic nuances in financial texts.

BERT's ability to process entire sequences of words simultaneously enables it to interpret complex financial language more effectively. Studies have demonstrated that transformer-based models outperform conventional approaches in sentiment classification, particularly when dealing with domain-specific texts such as financial news and earnings reports. This improvement in precision is crucial for extracting actionable insights from large-scale financial datasets.

Moreover, researchers have developed domain-specific variations of transformer models, such as FinBERT, which is fine-tuned on financial text corpora. These specialised models further enhance the accuracy of sentiment classification by incorporating industry-specific semantics and contextual nuances \citep{Araci2019}. By integrating such advanced NLP models with financial analytics, researchers can achieve a more precise understanding of investor sentiment and its implications for market behaviour.

BERT and its derivatives mark a paradigm shift in how sentiment is modelled, integrating pre-training on large corpora and fine-tuning on domain-specific tasks, making them highly adaptable \citep{schmidhuber2015,goodfellow2016}. These capabilities are crucial for real-time applications in financial forecasting and market monitoring.

\subsection{Geopolitical Events and Market Volatility}
Geopolitical conflicts are among the most powerful exogenous shocks to market behaviour \citep{hasan2022,mukhtarov2023}. Events such as wars, trade disputes and political crises introduce uncertainty into financial markets, prompting changes in investor sentiment and risk perception. \citep{baker2007} and \citep{bollen2011} demonstrated that market reactions to geopolitical instability often manifest themselves as increased volatility and abrupt price movements.

Studies such as \citep{smales2014} and \citep{drakos2010} demonstrate how geopolitical news sentiment, especially harmful content, triggers abrupt market reactions, often independently of underlying fundamentals. Media tone thus acts as both a transmitter and amplifier of volatility \citep{smales2014}. Social media platforms, such as Twitter (now X), further intensify this dynamic. The real-time dissemination of emotional responses and rumors on online platforms has been shown to cause rapid changes in trading behaviour and asset pricing \citep{bollen2011,sprenger2010}.

A notable case study in this context is the Russia-Ukraine conflict, which began on 22 February 2022. The onset of the conflict caused widespread uncertainty, leading to significant market fluctuations. Investors responded to news updates and geopolitical developments with increased sensitivity, contributing to rapid changes in market sentiment. The role of the media in shaping perceptions of the crisis further amplified these market reactions.

Understanding how geopolitical events influence market sentiment requires combining sentiment analysis techniques with financial models. By examining news articles, social media discussions and financial reports, researchers can assess the real-time impact of geopolitical events on investor sentiment and market dynamics. This approach provides valuable information for risk management and investment decision making.

\subsection{Application of GARCH Models in Volatility Analysis}

The Generalised Autoregressive Conditional Heteroscedasticity (GARCH) model has been widely employed in financial research to capture time-varying volatility patterns \citep{engle1982,bollerslev1986}. Introduced by \citep{engle1982}, the GARCH model enables researchers to analyse fluctuations in asset prices and assess the persistence of volatility shocks. The application of GARCH models is particularly relevant in periods of heightened uncertainty, such as geopolitical crises \citep{banerjee2024}.

Integrating sentiment analysis with GARCH models allows for a more comprehensive understanding of market dynamics. By incorporating sentiment-derived variables into volatility models, researchers can examine the extent to which sentiment fluctuations contribute to market instability. Studies have demonstrated that investor sentiment, as measured through financial news and social media, exhibits a strong correlation with market volatility, making it a valuable predictor of future price movements \citep{banerjee2024,tetlock2007}.

Recent advances in computational finance have further enhanced the effectiveness of sentiment-based volatility modelling. By leveraging machine learning techniques alongside econometric models, researchers can refine volatility forecasts and improve risk assessment methodologies \citep{pang2008,devlin2018}. This integration of sentiment analysis with GARCH models forms the foundation of our study, which explores the interplay between news sentiment and financial market volatility during the Russia-Ukraine conflict.

To improve robustness, researchers have introduced Student-t distributions in GARCH modeling to account for fat tails in financial data \citep{feng2017,lambert2001, bollerslev1987}. These enhancements are crucial for capturing the non-linear dynamics and extreme risks often present in geopolitical crises.

This literature review establishes the theoretical framework for our research, which applies the BERT model for sentiment extraction and GARCH models for volatility analysis. The subsequent sections detail the methodology, empirical findings, and implications of our study for financial market forecasting and risk management.

¡hecho! Aquí tienes **toda la metodología** con los **cambios de estructura** (secciones numeradas + labels, tokens en monoespacio, guion protegido en *Student-t*), **sin cambiar el contenido**.

```latex
\section{Methodology}\label{sec:method}
\subsection{Data}\label{sec:data}

This study explores the relationship between investor sentiment during geopolitical crises and financial market volatility by analysing U.S. news media coverage of the Russia-Ukraine conflict. To understand the financial market's reaction, news article headlines were collected from Goperigon, a platform that collects articles from the top 100 U.S. news sources. The data was compiled from January 1 to July 17, 2024, collecting over 10,000 headlines from major newspapers such as the New York Times, CNN and The Wall Street Journal, with emphasis on financial markets, politics, and global affairs.

The data collection process consisted of three steps. First, we stored metadata such as title, author, publication date, and URL in a database. Second, we searched the URLs for references to the war, focusing on relevant keywords such as "war," "Russia," and "Ukraine." Then we gathered all text items (p-tags) of the remaining URLs (title, author, date).

To ensure the accuracy of the headlines, all text was converted to lowercase, special characters and unnecessary punctuation were removed, and common misspellings were corrected. These steps are essential in reducing noise and improving the performance of downstream NLP models \citep{cambria2017}. The Python libraries utilised for this text clean-up included Natural Language Toolkit (NLTK), Regular Expression Operations (RE), and BeautifulSoup. Figure 1 illustrates the number of items collected:

\begin{figure}[!htbp]
  \centering
  \includegraphics[width=\linewidth]{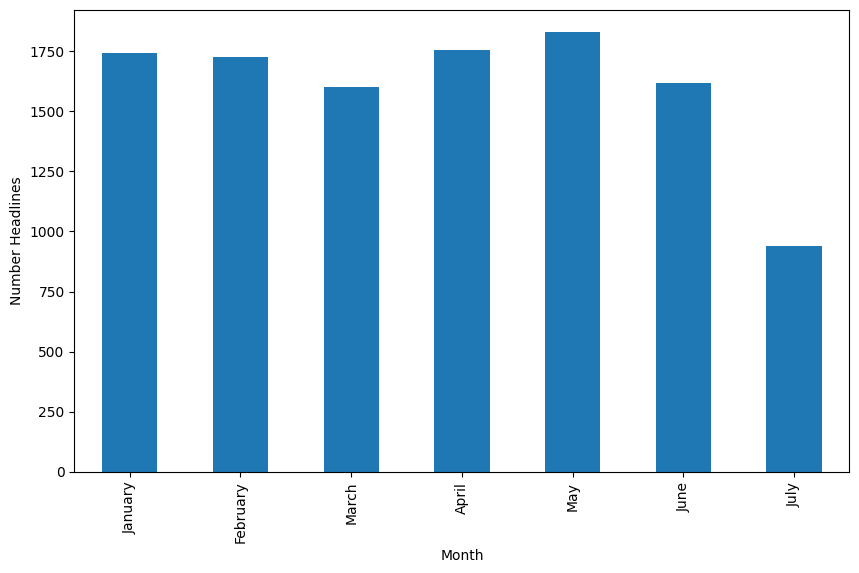} 
  \caption{Number of items collected over the sample period.}
  \label{fig:items_collected}
\end{figure}
\FloatBarrier

\subsection{Sentiment Analysis Model}\label{sec:sentiment}

BERT is an advanced NLP model introduced by \citet{devlin2018}. It utilises a transformer architecture that enables bidirectional understanding of the text, meaning the model considers the context of a word by analysing both the words before and after it in a sentence. This bidirectional processing allows BERT to capture nuanced meanings and complex relationships within the text, making it a powerful tool for various natural language understanding tasks \citep{zhang2018}.

This study investigates whether a negative sentiment about the Russia-Ukraine war, as reflected in U.S. news headlines,  is related to a decline in financial market performance. To achieve this goal, the BERT model is applied to classify the sentiment of the news headlines and generate a daily sentiment score. Using BERT in this context builds on prior research demonstrating the model’s effectiveness in financial and geopolitical contexts where language tends to be ambiguous and emotionally charged \citep{zhang2018}.

The initial stage requires separating a single sentence (news headlines) or two sentences together into a sequence of tokens. The first token in each sequence is a unique classification token \texttt{[CLS]}. This token's final hidden state is used as the aggregate sequence representation for classification tasks. Sentence pairs are clustered into a single sequence. As a result, we differentiate the sentences in two ways. First, we separate them with a special \texttt{[SEP]} token. Second, we add a learned embedding to each token indicating whether it belongs to phrases A or B. As shown in Figure 3, we denote the input embedding as E \citep{devlin2018}.

The input representation involves a comprehensive embedding system that includes token embeddings, segment embeddings, and position embeddings. A combination of these embeddings represents each token in the input text. Token embeddings provide the word's semantic meaning, segment embeddings distinguish between different segments (such as sentences) in the input, and position embeddings encode the position of each token in the sequence. This combination shown in Figure 2 allows the model to understand the context and positional relationships within the text \citep{devlin2018}. The sum of these types of embeddings is represented by: 

\[
E(x_i) = T(x_i) + S(x_i) + P(x_i)
\]

Where \(E(x_i)\) is the final embedding of token \(x_i\), \(T(x_i)\), \(S(x_i)\) and \(P(x_i)\) are the embeddings for token, segment and position, respectively. These embeddings are passed to the transformer layers, where multi-head attention computes contextual word representations \citep{vaswani2017}.

\begin{figure}[!htbp]
  \centering
  \includegraphics[width=\linewidth]{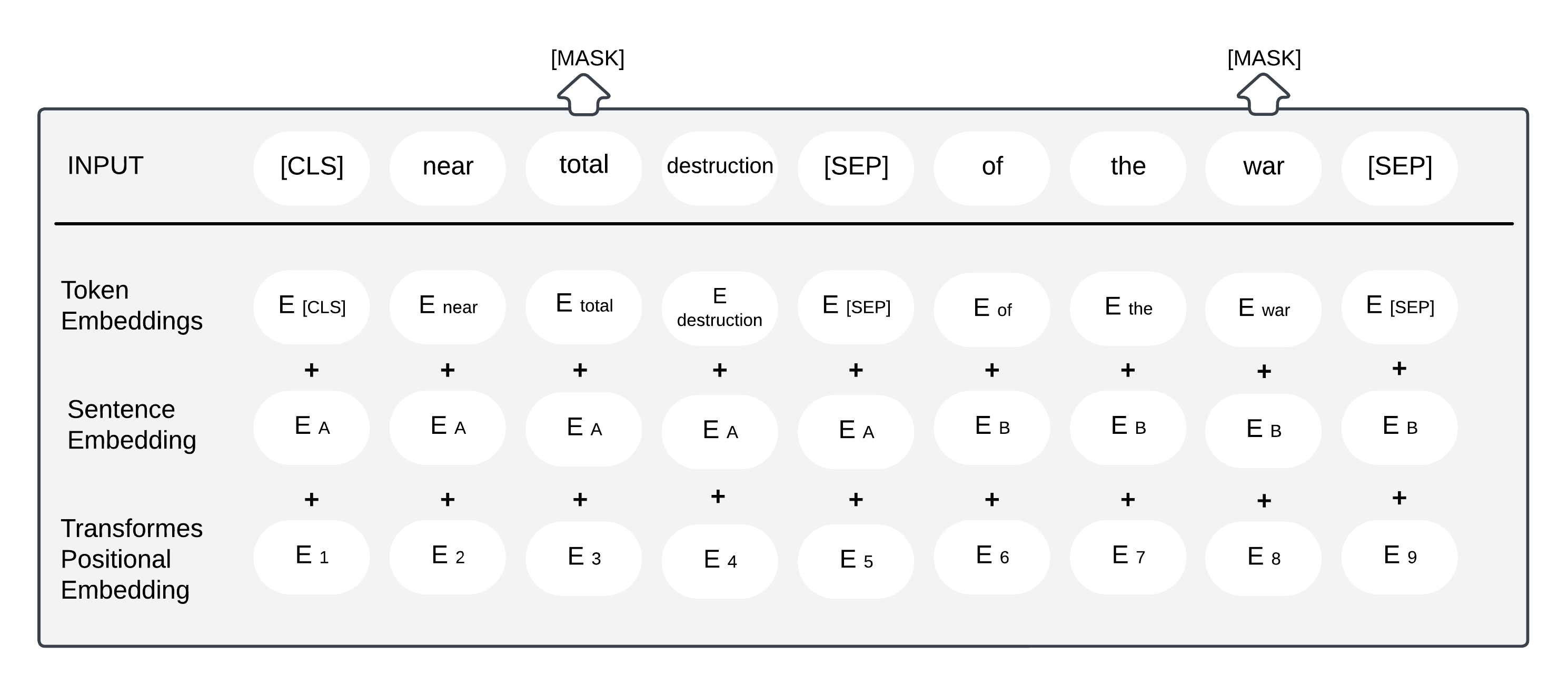} 
  \caption{Token, segment, and position embeddings composing the input representation \(E(x_i)\).}
  \label{fig:embeddings}
\end{figure}

Following the process described above, BERT initially pre-trained using a substantial corpus of unlabelled data by two principal tasks: Masked Language Modeling (MLM) and Next Sentence Prediction (NSP). MLM works by randomly masking a subset of the words in a sentence, requiring the model to predict these masked words based on the context of the other words, represented by the loss function:

\[
L_{\text{MLM}} = - \sum_{i \in M} \log P(x_i \mid x_{\text{masked}})
\]

Where \(M\) is the set of masked positions, and \(P\) represents the probability of predicting the original token \(x_i\) given the masked input. This task enables the model to learn deep word representations considering both preceding and succeeding contexts, which is essential for capturing complex meanings and relationships within the text \citep{devlin2018}.

NSP involves training the model to determine whether a given sentence  \(B\) logically follows another sentence  \(A\), enhancing its comprehension of sentence-level coherence: 

\[
L_{\text{NSP}} = -\log P(\text{IsNext} \mid A, B)
\]

Where "IsNext" is a binary label indicating if sentence \(B\) follows sentence \(A\). These processes, illustrated in Figure 3 are crucial in building a robust language understanding model \citep{devlin2018}.

\begin{figure}[!htbp]
  \centering
  \includegraphics[width=\linewidth]{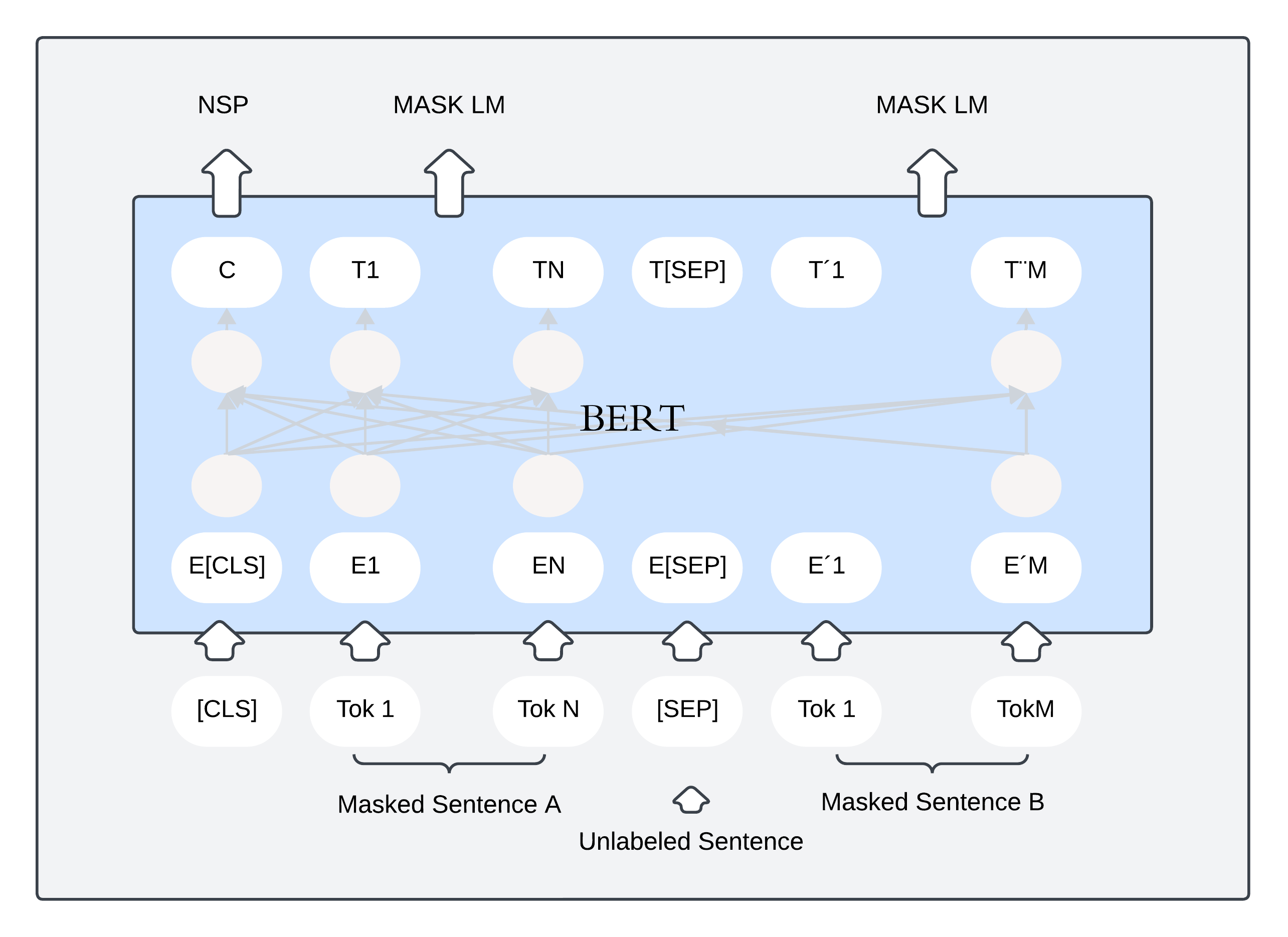} 
  \caption{BERT pre-training objectives: MLM and NSP.}
  \label{fig:pretrain}
\end{figure}
\FloatBarrier

After masking sentences to learn the underlying structure and meaning (pre-training), BERT can be fine-tuned for specific applications such as text classification, named entity recognition or question answering. Fine-tuning involves adjusting the model's pre-trained weights using a smaller task-specific dataset. The fine-tuned loss function is the following:

\[
L_{\text{fine-tune}} = - \sum_{i} y_i \log P(y_i \mid x_i ; \theta)
\]

Where \(y_i\) is the true label for the instance \(x_i\), and \(\theta\) represents the parameters of the model. The final output layer of the BERT model for sentiment analysis uses a softmax layer to classify the sentiments:

\[
\text{Sentiment} = \text{softmax}(W \cdot h_{\text{CLS}} + b)
\]

Where \(W\) y \(b\) are trainable parameters y \(h_{\text{CLS}}\) es el output del token \texttt{[CLS]}, usado como la representación agregada de toda la secuencia de entrada. 

The softmax function is commonly used in multi-class classification problems to convert raw model outputs (logits) into normalised probability distributions. Specifically, it ensures that each output lies between 0 and 1 and that the sum of all output values equals 1, allowing interpretation as class probabilities.

Formally, for a given input vector $z = (z_1, z_2, \dots, z_K)$ representing the non-normalised logit scores for $K$ classes, the softmax function is defined as:

\[
\text{softmax}(z_i) = \frac{e^{z_i}}{\sum_{j=1}^{K} e^{z_j}} \quad \text{for } i = 1, \dots, K
\]

This transformation emphasises the highest logit while suppressing the others, facilitating probabilistic interpretation. The predicted class is then obtained by selecting the index of the maximum softmax value, using the $\arg\max$ operator:

\[
\text{Predicted class} = \arg\max_i \text{softmax}(z_i)
\]

This step is crucial for classification tasks such as sentiment analysis, as it translates model output into interpretable predictions \citep{goodfellow2016}.
This flexibility is one of BERT's strengths, as it can be fine-tuned to perform exceptionally well across a wide range of NLP tasks as depicted in Figure 4, which shows the model being adapted for tasks such as MNLI, NER, and SQuAD \citep{devlin2018}.

\begin{figure}[!htbp]
  \centering
  \includegraphics[width=\linewidth]{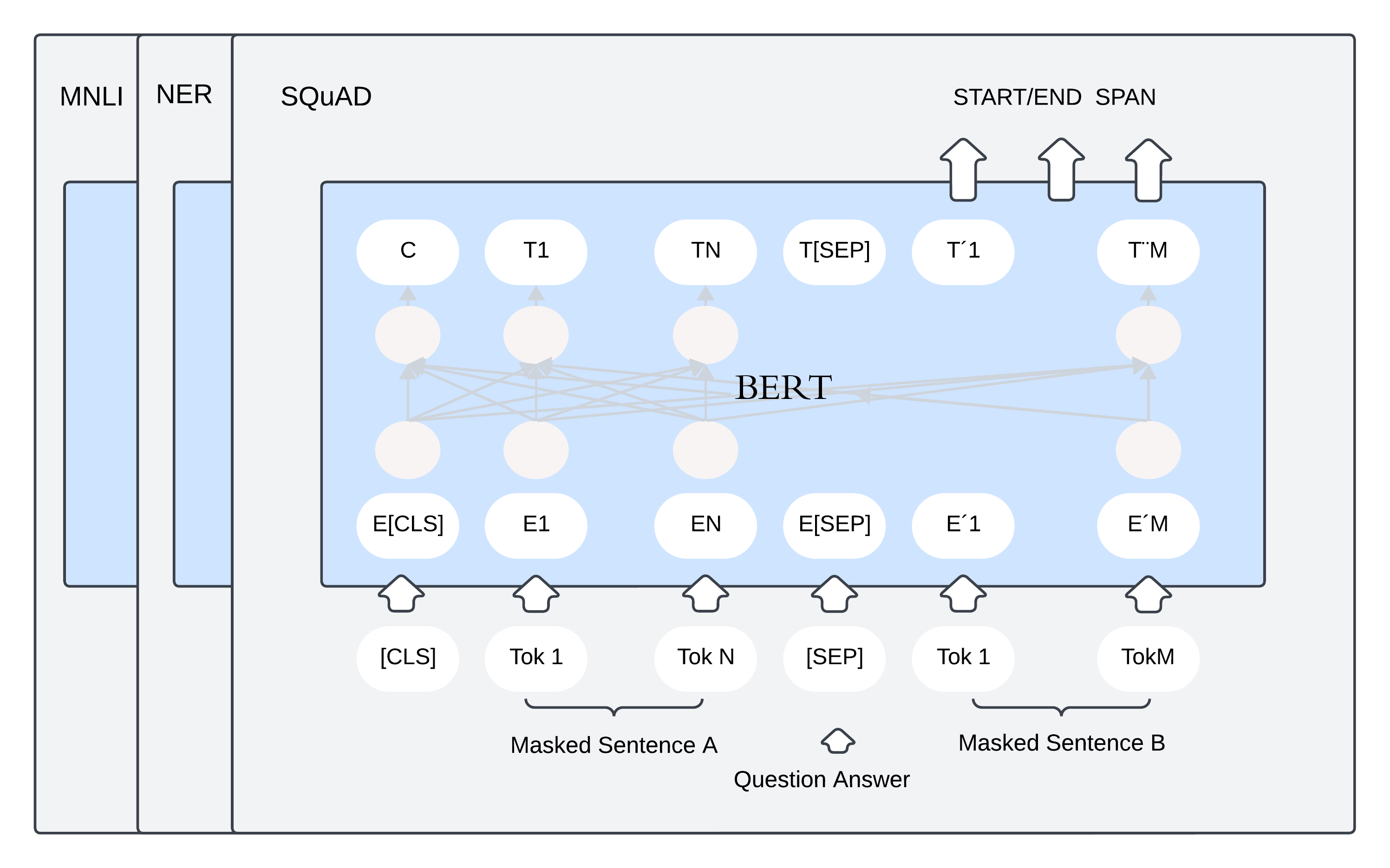} 
  \caption{High-level view of BERT fine-tuning for downstream NLP tasks.}
  \label{fig:finetune}
\end{figure}
\FloatBarrier

In the context of this study, the model was evaluated by dividing the data into 70\% for training, 15\% for testing, and 15\% for validation. This ensures a proper calibration of the model and a robust assessment of its performance. The BERT model classifies all tokenised sentences into sentiment categories: positive, negative, and neutral, and their respective probabilities known as logits. To calculate the sentiment, the following formula was used:

\[
\text{Sentiment} = \text{Logit}_{\text{positive sentiment}} - \text{Logit}_{\text{negative sentiment}}
\]

This calculation provides a quantitative measure of sentiment, reflecting the probability that the sentence is positive minus the probability that it is negative. This methodology follows similar approaches to the literature, such as that proposed by \citet{andriotis2014}. Relating detected sentiment changes to market dynamics influenced by geopolitical events is crucial. 

\subsection{Stock market and Russia-Ukraine war news sentiment}\label{sec:stock}

In this section, we explore the relationship between sentiment (as expressed in news headlines related to the Russia-Ukraine conflict) and financial markets. To quantify this relationship, an  Ordinary Least Squares (OLS) regression model was initially employed (Appendix 1). This approach serves as a baseline estimation to examine whether variations in daily sentiment scores can explain changes in stock market returns.
To ensure the robustness of the OLS estimation, a series of diagnostic tests were performed. These include calculation of Variance Inflation Factors (VIF) to assess multicollinearity among the explanatory variables, the Breusch-Pagan test to detect the presence of heteroscedasticity, and the Durbin-Watson test to evaluate autocorrelation in the model residuals. The results indicated heteroscedasticity in the error terms (Appendix 2),  which can lead to inefficient and biased estimators \citep{wooldridge2010}. 

To address heteroscedasticity and improve the reliability of the model, a Generalised Autoregressive Conditional Heteroscedasticity (GARCH) model was applied.  GARCH is a well-established financial time series modelling technique for conditional heteroscedasticity, ensuring more accurate and robust estimates \citep{bollerslev1986,francq2019}. In the context of financial data, it is crucial to model the time-varying volatility (heteroscedasticity) observed in asset returns, which the GARCH model captures by allowing the variance of the error terms to change over time.

For this study, GARCH recognises that the variance of errors in a regression model is not constant but varies over time, capturing the volatility that characterises financial returns. The model is based on the idea that the variance of the market returns is related to the past variance of the returns. In other words, if the market has been very volatile recently, the GARCH model expects it to remain volatile in the coming days. However, if the market has been stable in recent days, the model expects it to remain stable. GARCH improves the accuracy of the estimate by modelling the conditional volatility of the data, resulting in more efficient estimators.

Given the above, the following equation was proposed where the estimation method is considered:

\[
Y_t = \mu + \beta_1 \text{Sentiment}_t + \beta_2 \text{VIX}_t + \beta_3 \text{OFR}_t + \beta_4 \text{EPU}_t + \beta_5 \text{Bond}_t + \epsilon_t
\]

\[
\epsilon_t \sim N(0,\sigma_t^2)
\]

\[
\sigma_t^2 = \alpha_0 + \alpha_1 \epsilon_{t-1}^2 + \beta_1 \sigma_{t-1}^2
\]

Where \(Y_t\) represents the daily returns of the S\&P 500 index at time \(t\), \(\mu\) is the constant term, \(\text{Sentiment}_t\) is the average daily sentiment derived using the BERT model, \(\text{VIX}_t\), \(\text{OFR}_t\), \(\text{EPU}_t\), and \(\text{Bond}_t\) are the explanatory variables, \(\epsilon_t\) is the error term assumed to follow a normal distribution, and \(\sigma_t^2\) is the conditional variance which depends on past squared errors.

While the GARCH model assumes normally distributed residuals, many financial datasets exhibit fat tails, meaning the occurrence of extreme values is more frequent than expected under a normal distribution (Appendix 3).  To address this, we opted to model the residuals using the Student\mbox{-}t distribution, which accounts for the heavier tails commonly seen in financial returns \citep{feng2017}. The Student\mbox{-}t distribution introduces  degrees of freedom that help to adjust the tail heaviness, improving the model’s robustness in estimating volatility risks \citep{bollerslev1986, lambert2001}. By doing so, GARCH with a Student\mbox{-}t distribution enhances accuracy in volatile markets. The model can be written as: 
\[
\sigma_t^2 = \alpha_0 + \alpha_1 \epsilon_{t-1}^2 + \beta_1 \sigma_{t-1}^2
\]

Where \(\sigma_t^2\) is the conditional variance, \(\epsilon_{t-1}^2\) is the lagged squared residual, and \(\sigma_{t-1}^2\) is the lagged variance. The residuals \(\epsilon_t\) follow a Student\mbox{-}t distribution:

\[
f(\epsilon_t) = \frac{\Gamma\left(\frac{v+1}{2}\right)}{\Gamma\left(\frac{v}{2}\right) \sqrt{\pi(v-2)\sigma^2}} \left( 1 + \frac{\epsilon_t^2}{(v-2)\sigma^2} \right)^{-\frac{v+1}{2}}
\]

Where \(v\) represents the degrees of freedom, determining the heaviness of the tails. Lower values of \(v\) capture heavier tails, while higher values approach a normal distribution. This modification allows the model to more effectively capture tail risk, which is critical for analysing extreme market movements common in financial datasets.

Furthermore, to calculate the daily returns of the S\&P 500 index, the logarithmic returns formula was applied:

\[
R_t = \ln \left(\frac{P_t}{P_{t-1}}\right)
\]

Where \(R_t\) represents the return of the S\&P 500 at time \(t\), \(P_t\) is the closing price at \(t\), and \(P_{t-1}\) is the closing price at \(t-1\). This calculation captures the percentage change in the index from one day to the next, enabling the analysis of volatility and trends in returns over the studied period. On the other hand, for the explanatory variables at time \(t\), we include the Volatility Index (VIX), Financial Stress Index (OFR), Economic Policy Uncertainty Index (EPU), and the U.S. 10-year bond yield (Bond). Historical data for these variables were obtained for the same period as the online articles for the BERT sentiment analysis. Spline interpolation was applied to ensure the completeness and continuity of this financial data. This technique was employed to prevent data loss and ensure a more robust analysis by maintaining the integrity of the time series. To calculate the spline interpolation, the following formula was applied:

\[
S(x) = \sum_{i=1}^{n} a_i B_i(x)
\]

Where  \(B_i(x)\) are the basis functions that define the shape of the spline between control points. These basis functions are typically piecewise polynomial functions constructed to ensure smooth transitions and continuity at the knots. Spline interpolation was applied to ensure that all required data points were available for subsequent analysis. This method provided a smooth and continuous approximation across missing or irregular time intervals, offering a reliable foundation for examining the relationship between geopolitical events and market reactions \citep{deboor2001}. Prior to analysis, the statistical properties of each variable were evaluated to understand their distribution and variability, ensuring the robustness of the methodological approach.
\section{Results}\label{sec:results}

This section presents the empirical results examining the relationship between news sentiment surrounding the Russia-Ukraine conflict and market volatility in the U.S. stock market. Using the BERT sentiment scores and a GARCH model specification with a Student-t distribution, we tested whether a negative sentiment correlates with greater financial market instability. Figure 5 presents the average daily sentiment score derived from U.S. media headlines over the sample period (January 1 – July 17, 2024). The sentiment values generally exhibit a predominantly negative trajectory, consistent with ongoing geopolitical tension. The sentiment score is calculated as the difference between the logits of positive and negative classification outputs from the fine-tuned BERT model.

\begin{figure}[!htbp]
  \centering
  \includegraphics[width=\linewidth]{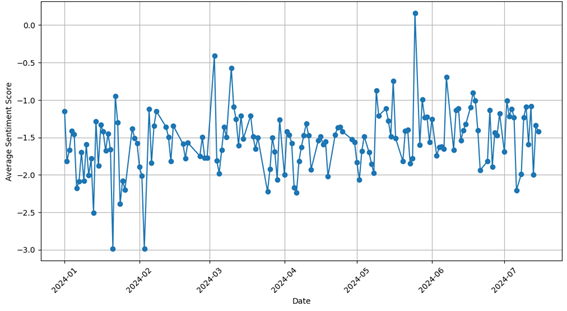} 
  \caption{Average daily sentiment score derived from U.S. media headlines (Jan 1–Jul 17, 2024).}
  \label{fig:sentiment_daily}
\end{figure}
\FloatBarrier

Figure 6 shows the distribution of sentiment categories across the dataset. The majority of headlines fall into the negative or neutral category, reinforcing the nature of the conflict as perceived by major U.S. media outlets.

\begin{figure}[!htbp]
  \centering
  \includegraphics[width=\linewidth]{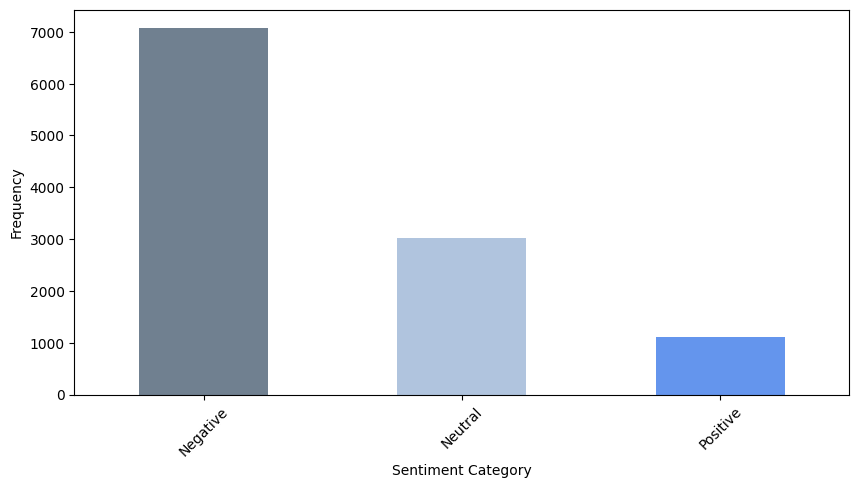} 
  \caption{Distribution of sentiment categories across the sample.}
  \label{fig:sentiment_distribution}
\end{figure}
\FloatBarrier

The sentiment scores, along with a set of control variables (VIX, 10-year bond yield, Financial Stress Index, and Economic Policy Uncertainty Index), were regressed against S\&P 500 daily returns using a GARCH(1,1) model. GARCH enables the modelling of time-varying volatility, a critical characteristic of financial returns during periods of geopolitical uncertainty.
Table 1 presents the estimation results. The coefficient for the Sentiment Score is -0.2275, statistically significant at the 1\% level (p = 0.0016). This indicates that negative news sentiment is significantly associated with increased stock market volatility. These findings support the hypothesis that investor pessimism driven by media coverage contributes to elevated uncertainty in financial markets.
The VIX, widely recognised as a proxy for investor fear and forward-looking market volatility, also shows a significant negative relationship (coefficient = -0.2865, p = 0.0094). This confirms that periods with heightened expected volatility (as reflected in the VIX) correspond to higher realised volatility in the market.
By contrast, the remaining control variables, 10-year bond yield, Financial Stress Index (OFR), and Economic Policy Uncertainty Index (EPU), were not statistically significant. These results suggest that, in the presence of media-driven sentiment and VIX, traditional macroeconomic indicators offer limited additional explanatory power for short-term market volatility during geopolitical crises. 

\begin{table}[!htbp]
\centering
\caption{Coefficient Estimates from GARCH (1,1) Model for S\&P 500 returns, incorporating News Sentiment and Control Variables.}
\label{tab:garch_results}
\begin{tabular}{lccc}
\hline
\textbf{Variable} & \textbf{Coefficient} & \textbf{Standard Error} & \textbf{p-value} \\
\hline
Constant        & -0.2545 & 2.2398 & 0.9098 \\
Sentiment Score & -0.2275 & 0.0703 & 0.0016 \\
VIX             & -0.2865 & 0.1083 & 0.0094 \\
Bond 10-years   &  0.0016 & 0.0008 & 0.2022 \\
OFR             &  0.3966 & 0.5022 & 0.4316 \\
EPU             & -0.0012 & 0.0010 & 0.3249 \\
\hline
Observations    & \multicolumn{3}{c}{105} \\
Adjusted R$^2$  & \multicolumn{3}{c}{0.1481} \\
F-statistic     & \multicolumn{3}{c}{4.687} \\
p-value         & \multicolumn{3}{c}{0.0006} \\
\hline
\multicolumn{4}{p{12cm}}{\small Note: p-values reflect the probability of the coefficients being zero. Significance: $p<0.001$, $p<0.01$, $p<0.05$, $p<0.1$.}
\end{tabular}
\end{table}
\FloatBarrier

\section{Discussion}\label{sec:discussion}

The results obtained in this study align with previous findings in the literature that explore the relationship between geopolitical events and financial markets. \citep{banerjee2024} demonstrate that geopolitical risks heighten investor uncertainty and elevate perceived market risk, leading to pronounced volatility. Their analysis of international markets under geopolitical stress found that negative news sentiment contributes to increased price instability, particularly during periods of escalating conflict. This is consistent with the present study's GARCH results, which show that the Sentiment Score derived using BERT is significantly and negatively associated with stock market volatility, reinforcing that investor psychology, shaped by media narratives, is a crucial determinant of short-term market dynamics during crises.

\citep{smales2014} also emphasises the important role of sentiment during geopolitical conflicts. Their work shows that investors react more sharply to negative sentiment when uncertainty is high, mainly when such sentiment is from trusted news sources. These observations align closely with our findings, where negative sentiment extracted from major U.S. news headlines showed a statistically significant effect on volatility, independent of economic fundamentals. This confirms that sentiment extracted via advanced NLP models can be an early warning indicator of financial turbulence, particularly in conflict-driven markets.

The results also confirm the important role of the VIX as a predictor of market volatility. As \citep{shang2019} highlighted, the VIX serves as both a forward-looking gauge of market expectations and a proxy for investor fear. In the present study, the VIX coefficient was statistically significant and negative, reinforcing its utility as a benchmark for market anxiety. By including both media-based sentiment and the VIX in the volatility model, this framework captures behavioural and anticipatory dimensions of investor perception. This dual mechanism enhances the model’s explanatory power, particularly in environments where rational expectations frameworks fail to reflect the full extent of market stress.

By contrast, traditional financial indicators such as bond yields, the Financial Stress Index (OFR), and the Economic Policy Uncertainty Index (EPU) were not found to be statistically significant. These results suggest that sentiment-based and forward-looking measures dominate over macroeconomic fundamentals in explaining volatility during periods of geopolitical tension \citep{bloom2009}. This observation aligns with \citep{yilmazkuday2024}, who argue that geopolitical shocks introduce non-linear, sentiment-driven responses that conventional indicators cannot adequately capture. As uncertainty becomes narrative-based rather than strictly data-driven, traditional indicators may lose predictive power in the short term.

The empirical performance of the GARCH model with a Student-t distribution further supports the view that financial returns during crises are prone to heavy tails and extreme deviations. \citep{bollerslev1987, francq2019}  point out that financial markets are characterised by volatility clustering and non-Gaussian return distributions, particularly in response to exogenous shocks. By accounting for excess kurtosis and tail risk, the Student-t specification enhances model robustness, a feature supported by \citep{feng2017}, who show that this adjustment significantly improves volatility forecasts under crisis conditions. The model used in this study captures these statistical irregularities, reflecting a more realistic representation of financial market behaviour under stress.

Importantly, the findings reinforce the growing importance of AI-driven sentiment analysis in financial research and forecasting. Traditional sentiment tools such as dictionary-based methods \citep{loughran2011} or survey-based indicators often struggle with context, sarcasm, or evolving language use. In contrast, as a transformer-based model, BERT captures semantic nuance, word order, and bidirectional context, making it  well-suited for sentiment extraction from complex geopolitical narratives \citep{devlin2018,Araci2019}. The effectiveness of BERT in this study adds to the evidence from \citep{zhang2018, gupta2020}  who found that transformer models outperform SVMs and recurrent networks in domain-specific text tasks, including finance.

This study also speaks to a broader behavioural shift in financial markets. The increasing sensitivity of asset prices to media sentiment and narrative framing suggests a trend toward sentiment-driven trading, where market behaviour is guided less by fundamentals and more by real-time emotional signals \citep{tetlock2007, schmeling2009}. This behavioural response is consistent with the theory of limited attention and bounded rationality, where investors, overwhelmed by information, rely on emotionally salient cues to make decisions \citep{barberis2018}. The persistence of negative sentiment over the sample period and its predictive link to volatility offer strong evidence that financial markets are becoming more psychologically responsive, particularly in times of global tension.

Additionally, the absence of explanatory power in the EPU index during this crisis highlights the growing mismatch between top-down macroeconomic indicators and bottom-up media narratives. As \citep{ahern2014} observe, the tone of political news, not just the content, can significantly affect investor decisions. This divergence implies that quantitative models integrating AI-based textual analysis with market data can provide superior forecasting power compared to traditional econometric models that exclude unstructured information.
Moreover, these insights carry clear, practical implications. Investment managers and risk officers can leverage AI-driven sentiment metrics to improve volatility forecasting and hedging strategies. Analysts can better detect inflexion points in investor sentiment by integrating BERT sentiment signals into dashboard-based monitoring tools and calibrating their exposure accordingly. This approach is particularly valuable during crises when price movements are driven by uncertainty, expectation, and narrative rather than earnings reports or macroeconomic data \citep{smales2014}.
The study offers policymakers a novel tool for assessing market sentiment and stability. In real-time crisis response, regulatory institutions can monitor sentiment-based indicators to gauge public perception and its effects on market liquidity and volatility. As financial markets become more intertwined with digital news flows and investor psychology, understanding media sentiment becomes vital to managing systemic risk.
To conclude, this study's findings support the view that financial markets are increasingly responsive to geopolitical developments and media narratives, reflecting a departure from models that assume rational behaviour under complete information. The significant impact of sentiment on volatility, the robustness of the Student-t GARCH model, and the limitations of traditional indicators underscore the value of combining AI-based sentiment analysis with econometric modelling to better understand and predict market behaviour in uncertain times.

\section{Conclusions and Recommendations}\label{sec:conclusions}

This study provides empirical evidence that advances our understanding of the relationship between media sentiment and financial market volatility during geopolitical crises. By integrating NLP techniques like the BERT transformer model with a Student-t GARCH econometric framework, the research demonstrates that sentiment extracted from news headlines can significantly predict stock market volatility. The results show that negative sentiment regarding the Russia-Ukraine conflict, as quantified through BERT, is strongly associated with increased volatility in the S\&P 500 index.

Including the Volatility Index (VIX) as a control variable further reinforces its role as a robust predictor of financial market stress, consistent with prior studies on volatility \citep{shang2019,banerjee2024}. In contrast, more traditional macroeconomic indicators, such as bond yields, the Financial Stress Index, and the Economic Policy Uncertainty Index, were not found to be significant, suggesting that during crises, market behaviour is impacted by real-time sentiment shifts and investor expectations, rather than by slow-moving economic fundamentals \citep{yilmazkuday2024,smales2014}.

This paper contributes to the growing literature on AI applications in finance, providing a framework for integrating deep learning-based sentiment extraction with time-series volatility models. It highlights the importance of using transformer architectures like BERT to analyse specific text, particularly during periods of information overload and heightened uncertainty when traditional forecasting tools may underperform \citep{devlin2018, zhang2018}. The findings also underscore the behavioural transformation of financial markets, where media narratives and psychological responses increasingly influence asset prices and risk dynamics \citep{tetlock2007,barberis2018}.

Despite its contributions, the study has limitations that open avenues for further research. First, the sentiment model relies solely on news sources from the U.S., which may introduce a geographic or cultural bias in the sentiment signals. Future research could expand the dataset to include international and multilingual news outlets, enabling a more comprehensive global sentiment analysis and cross-market spillovers.

Second, the analysis operates on daily data, which may mask intraday sentiment swings that affect high-frequency trading and liquidity \citep{bollen2011,smales2014}. Future studies could incorporate higher-frequency sentiment analysis, using timestamped headlines and real-time news feeds to explore sentiment volatility at hourly or minute-level granularity.

Third, while this study applies the general BERT model, emerging transformer-based architectures such as FinBERT \citep{Araci2019}, RoBERTa \citep{liu2020}, and DeBERTa \citep{he2021} have shown superior performance in financial and domain-adapted text classification tasks. A comparative evaluation of these models within the same GARCH volatility framework could provide additional insight into model selection and optimal design for sentiment-driven forecasting.

Fourth, future work could explore sectoral or firm applications, examining whether sentiment impacts specific industries (e.g. energy, defence, technology) differently during geopolitical crises. This sector-level perspective would enhance understanding of heterogeneous investor responses and the cross-sectional implications of sentiment-driven trading behaviour \citep{ahern2014}.

Finally, this research invites exploration into the development of automated dashboards and forecasting platforms that fuse sentiment data, market analytics, and risk signals in real-time. Such systems could serve as decision-support tools for portfolio managers, traders, and regulators, providing an integrated view of behavioural risk, market structure, and macro-political dynamics \citep{Gu2020, smales2014}.

To conclude, this study reinforces the value of combining AI-based sentiment models with econometric volatility modelling to enhance understanding of financial market behaviour during crises. As geopolitical uncertainty and information intensity continue to shape global markets, sentiment analysis rooted in deep learning offers a powerful tool for navigating and forecasting volatility in real-time. This intersection of applied artificial intelligence, behavioural finance, and risk analytics marks a promising direction for future academic inquiry and practical implementation.


\section*{Data Availability}
The data and code supporting the findings of this study are publicly available at the following GitHub repository:  

\url{https://github.com/DomSop/geopolitical-sentiment-volatility}

The repository includes the cleaned sentiment dataset, the Python and R scripts used for analysis, and instructions for replicating the results. The original raw news data was obtained from Goperigon (\url{https://goperigon.com/}) under a non-commercial research license and cannot be redistributed. However, the transformed dataset used in the analysis is fully accessible.

\section*{Declaration of Interest Statement}
The authors declare no conflict of interest.

\clearpage
\appendix
\section{Appendices}\label{sec:appendix}

\subsection{OLS Model}\label{app:ols}

\begin{table}[!htbp]
\centering
\caption{Estimates of S\&P 500 returns using Russia–Ukraine war news sentiments plus a set of controls.}
\label{tab:ols_results}
\begin{tabular}{lccc}
\toprule
\textbf{Variable} & \textbf{Coefficient} & \textbf{Standard Error} & \textbf{p-value} \\
\midrule
Constant          & -0.2618 & 2.1996 & 0.9055 \\ 
Sentiment Score   & -0.2310 & 0.0690 & 0.0011 \\ 
VIX               & -0.3059 & 0.1063 & 0.0049 \\ 
Bond 10-years     &  0.0011 & 0.0008 & 0.1666 \\ 
OFR               &  0.4033 & 0.4932 & 0.4154 \\ 
EPU               & -0.0009 & 0.0010 & 0.3593 \\ 
\midrule
Observations      & \multicolumn{3}{c}{105} \\ 
Adjusted R$^2$    & \multicolumn{3}{c}{0.166} \\ 
F-statistic       & \multicolumn{3}{c}{5.216} \\ 
p-value           & \multicolumn{3}{c}{0.0003} \\ 
\bottomrule
\multicolumn{4}{p{12cm}}{\small Note: p-values reflect the probability of the coefficients being zero. Significance: \(p<0.001\), \(p<0.01\), \(p<0.05\), \(p<0.1\).}
\end{tabular}
\end{table}
\FloatBarrier

\subsection{Robustness Tests}\label{app:robustness}

\begin{table}[!htbp]
\centering
\caption{Breusch–Pagan Test Results}
\label{tab:bp}
\begin{tabular}{lcc}
\toprule
\textbf{Statistic} & \textbf{Value} & \textbf{p-value} \\
\midrule
BP Statistic       & 15.213 & 0.0095 \\
Degrees of Freedom & 5      &   –    \\
\bottomrule
\multicolumn{3}{p{12cm}}{\small Notes: \(p<0.01\) indicates significant heteroscedasticity.}
\end{tabular}
\end{table}

\begin{table}[!htbp]
\centering
\caption{Durbin–Watson Test Results}
\label{tab:dw}
\begin{tabular}{lcc}
\toprule
\textbf{Statistic} & \textbf{Value} & \textbf{p-value} \\
\midrule
DW Statistic & 1.6433 & 0.0164 \\
\bottomrule
\multicolumn{3}{p{12cm}}{\small Notes: \(p<0.05\) indicates significant autocorrelation.}
\end{tabular}
\end{table}

\begin{table}[!htbp]
\centering
\caption{Variance Inflation Factor (VIF) Results}
\label{tab:vif}
\begin{tabular}{lc}
\toprule
\textbf{Variable} & \textbf{VIF} \\
\midrule
Sentiment Score & 1.066 \\
VIX             & 2.955 \\
Bond 10-years   & 3.170 \\
OFR FSI         & 2.848 \\
EPU             & 1.134 \\
\bottomrule
\multicolumn{2}{p{12cm}}{\small Notes: VIF \(> 10\) suggests significant multicollinearity.}
\end{tabular}
\end{table}

\begin{figure}[!htbp]
\centering
\includegraphics[width=\textwidth]{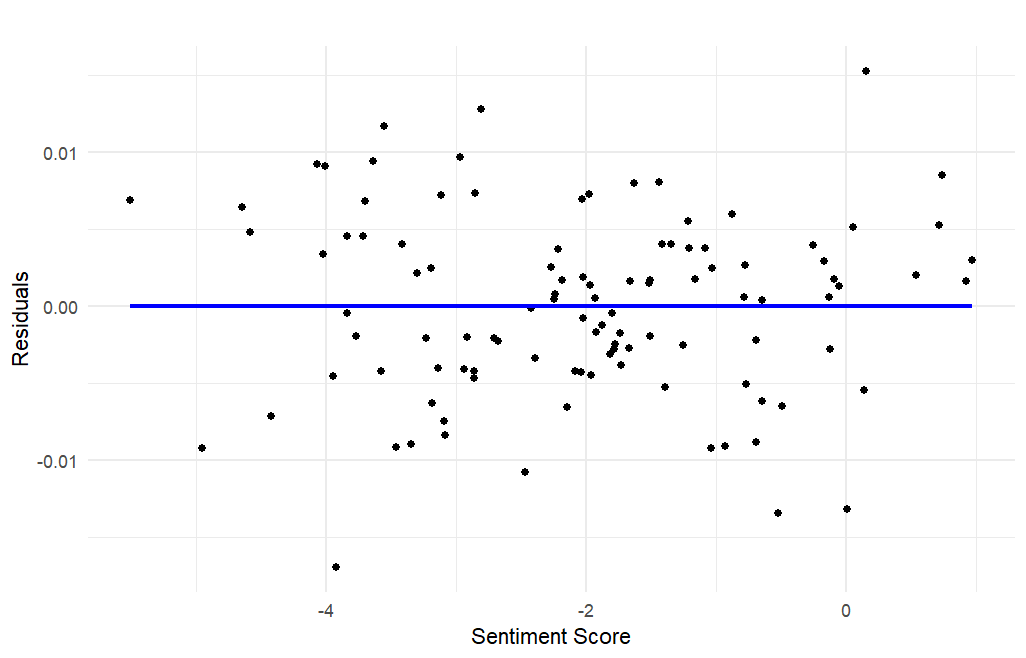}
\caption{Residuals vs. Sentiment Score.}
\label{fig:residuals_vs_sentiment}
\end{figure}
\FloatBarrier

\subsection{Q–Q Plot of GARCH Residuals}\label{app:qq}

\begin{figure}[!htbp]
\centering
\includegraphics[width=\textwidth]{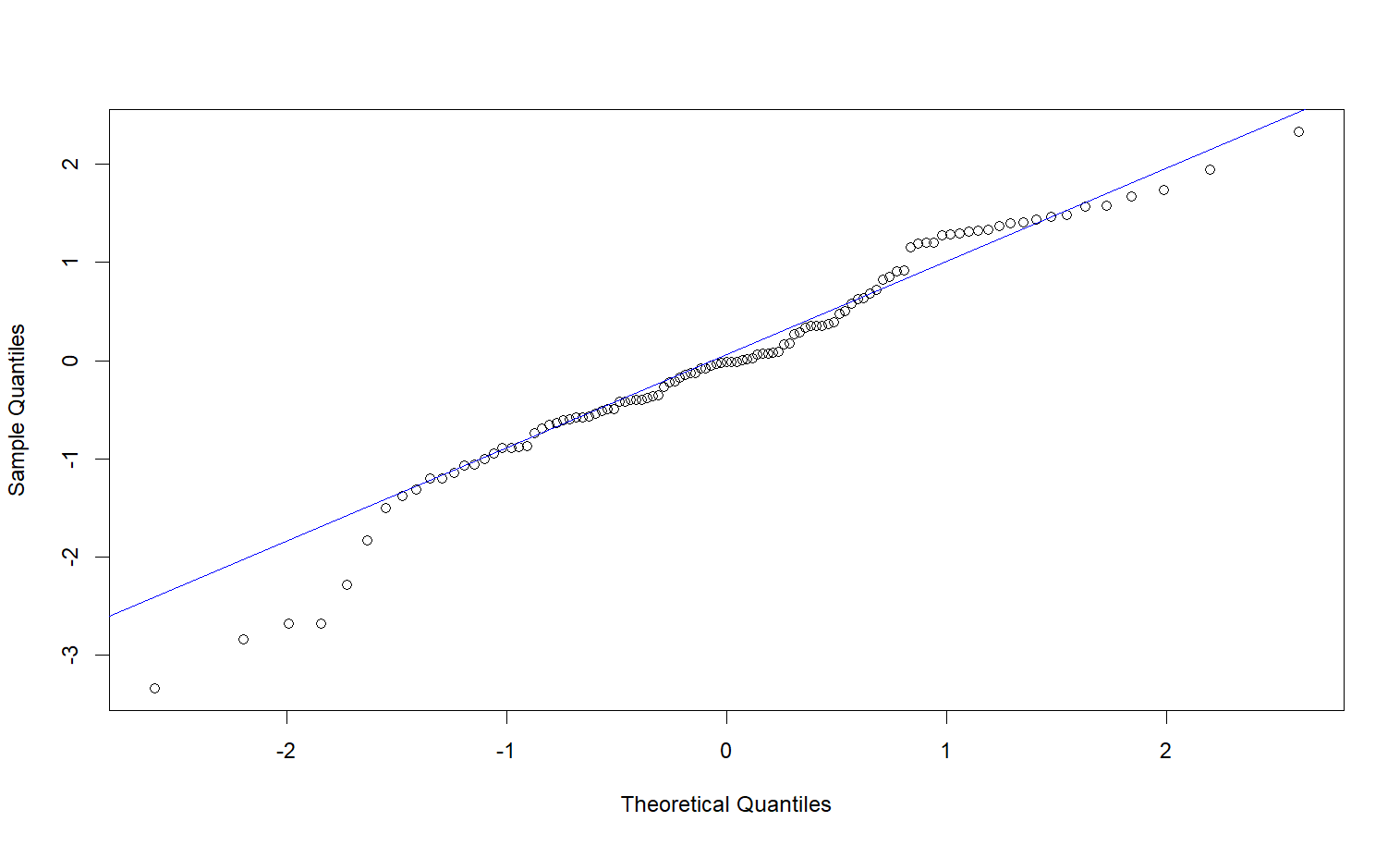}
\caption{Q–Q plot for checking normality of GARCH residuals.}
\label{fig:qq_garch}
\end{figure}
\FloatBarrier

\clearpage
\bibliographystyle{unsrtnat}  

\bibliography{references}      
\end{document}